\documentclass[aps,prb,twocolumn,showpacs,superscriptaddress]{revtex4}
\usepackage{amsmath}
\usepackage{amssymb}
\usepackage{epsfig}
\usepackage{color}
\usepackage{graphicx,amsmath}

\begin{document}
\title{Challenging Magnetic Field Dependence of the Residual Resistivity
of the Heavy-Fermion Metal CeCoIn$_5$}
\author{V. R. Shaginyan}\email{vrshag@thd.pnpi.spb.ru}
\affiliation{Petersburg Nuclear Physics Institute, Gatchina,
188300, Russia}\affiliation{Clark Atlanta University, Atlanta, GA
30314, USA} \author{A. Z. Msezane}\affiliation{Clark Atlanta
University, Atlanta, GA 30314, USA}\author{K. G.
Popov}\affiliation{Komi Science Center, Ural Division, RAS,
Syktyvkar, 167982, Russia}
\author{J.~W.~Clark}
\affiliation{McDonnell Center for the Space Sciences \&
Department of Physics, Washington University,
St.~Louis, MO 63130, USA}
\author{M.~V.~Zverev}
\affiliation{Russian Research Centre Kurchatov
Institute, Moscow, 123182, Russia}
\affiliation{Moscow Institute of Physics and Technology, Moscow, 123098, Russia}
\author{V. A. Khodel} \affiliation{Russian Research Centre Kurchatov Institute,
Moscow, 123182, Russia} \affiliation{McDonnell Center for the Space
Sciences \& Department of Physics, Washington University, St.~Louis,
MO 63130, USA}

\begin{abstract}
An explanation of paradoxical behavior of the residual resistivity
$\rho_0$ of the heavy-fermion metal CeCoIn$_5$ in magnetic fields
and under pressure is developed.  The source of this behavior is
identified as a flattening of the single-particle spectrum, which
exerts profound effects on the specific heat, thermal expansion
coefficient, and magnetic susceptibility in the normal state, the
specific heat jump at the point of superconducting phase
transition, and other properties of strongly correlated electron
systems in solids.  It is shown that application of a magnetic
field or pressure to a system possessing a flat band leads to a
strong suppression of $\rho_0$.  Analysis of its measured
thermodynamic and transport properties yields direct evidence for
the presence of a flat band in CeCoIn$_5$.
\end{abstract}

 \maketitle

Measurements of the resistivity $\rho(T)$ in external magnetic fields $H$ have
revealed a diversity of low-temperature behaviors of this basic property in
heavy-fermion (HF) metals, ranging from the familiar Landau Fermi liquid (LFL)
character to challenging non-Fermi liquid (NFL) behavior \cite{pag1,ron,pag2,loh}.
The resistivity $\rho(T)$ is frequently approximated by the formula
\begin{equation}
\rho(T)=\rho_0+AT^n,\label{res}
\end{equation}
where $\rho_0$ is the residual resistivity and $A$ is a
$T$-independent coefficient. The index $n$ takes the values 2 and
1, respectively, for FL and NFL behaviors and $1\lesssim n\lesssim
2$ in the crossover between.  The term $\rho_0$ is ordinarily
attributed to impurity scattering.  The application of the weak
magnetic field is known to produce a positive classical
contribution $\propto H^2$ to $\rho$ arising from orbital motion of
carriers induced by the Lorentz force.  However, when considering
spin-orbit coupling in disordered electron systems where electron
motion is diffusive, the magnetoresistivity may have both positive
(weak localization) and negative (weak anti-localization) signs
\cite{larkin}.

\begin{figure}[!ht]
\begin{center}
\includegraphics [width=0.47\textwidth]{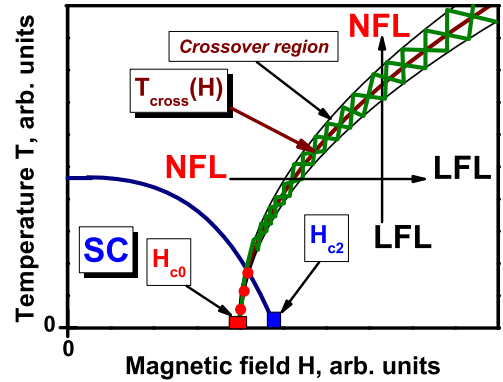}
\end{center}
\caption{(color online). Schematic $T-H$ phase diagram of
CeCoIn$_5$. The vertical and horizontal arrows crossing the
transition region marked by the thick jagged lines depict the
LFL-NFL and NFL-LFL transitions at fixed $H$ and $T$, respectively.
As shown by the solid curve, at $H<H_{c2}$ the system is in its
superconducting (SC) state, with $H_{c0}$ denoting a quantum
critical point hidden beneath the SC dome where the flat band could
exist at $H\leq H_{c0}$. The hatched area with the solid curve
$T_{\rm cross}(H)$ represents the crossover separating the domain
of NFL behavior from the LFL domain. A part of the crossover marked
with the dots is hidden in the SC state. The NFL state is
characterized by the entropy excess $S_{*}$ of Eq.~\eqref{S*}.}
\label{fig1}
\end{figure}

Our focus here is on the compound CeCoIn$_5$, whose $H-T$ phase diagram is
drawn schematically in Fig.~\ref{fig1}.  Its resistivity $\rho(T,H)$ differs
from zero in the region beyond the solid curve that separates the superconducting
(SC) and normal states.  Above the critical temperature $T_c$ of the SC phase
transition, the zero-field resistivity $\rho(T,H=0)$ varies linearly with $T$.
On the other hand, at $T\to 0$ and magnetic fields $H\geq H_{c2}\simeq5$ T,
the curve $\rho(T,H_{c2})$ is parabolic in shape \cite{pag1,pag2}.
As studied experimentally, CeCoIn$_5$ is one of the purest heavy-fermion metals.
Hence the applicable regime of electron motion is ballistic rather than diffusive,
and both weak and anti-weak localization scenarios are irrelevant.  Accordingly,
one expects the $H$-dependent correction to $\rho_0$ to be positive and small.
Yet this is far from the case: specifically $\rho_0(H=0)\simeq1.5$ ${\rm \mu\Omega cm}$,
while $\rho_0(H=6\,{\rm T})\simeq 0.3$ ${\rm \mu\Omega cm}$ \cite{pag1,pag2}.

To resolve this paradox, we suggest that the electron system of
CeCoIn$_5$ contains a flat band.  Flattening of the single-particle
spectrum $\epsilon({\bf p})$ is directly relevant to the problem
addressed since, due to Umklapp processes, quasiparticles of the
flat band produce a contribution to $\rho_0$ indistinguishable from
that due to impurity scattering \cite{kz,kzc}.  Furthermore, it is
crucial that the flat band somehow becomes depleted at $T\to 0$ and
$H=6$ T, to avoid contradiction of the Nernst theorem.  This
depletion entails a dramatic suppression of the flat-band
contribution to $\rho_0$.

Before proceeding to the analysis of this suppression and its
relevance to the behavior observed in CeCoIn$_5$, we call attention
to salient aspects and consequences of the flattening of
$\epsilon({\bf p})$ in strongly correlated Fermi systems. The
theoretical possibility of this phenomenon, also called swelling of
the Fermi surface or fermion condensation, was discovered two
decades ago \cite{ks,vol,noz} (for recent reviews, see
\cite{shagrep,shag,mig100}).  It has received new life in the
conceptual framework of {\it topological matter}, where it is
characterized by nontrivial topology of the Green function in
momentum space and associated with topologically protected flat
bands \cite{volHe,volv,lee,green,voltop}. At $T=0$, the ground
state of a system having a flat band is degenerate, and therefore
the occupation numbers $n_*({\bf p})$ of single-particle states
belonging to the flat band, which form a so-called fermion
condensate (FC), are continuous functions of momentum that
interpolate between the standard LFL values $\{0,1\}$.  This leads
to an entropy excess
\begin{equation}
S_*=-\sum_{\,{\bf p}} n_*({\bf p})\ln n_*({\bf p})+(1-n_*({\bf
p}))\ln(1- n_*({\bf p})),
\label{S*}
\end{equation}
which does not contribute to the specific heat $C(T)$.  However, in
contrast to the corresponding LFL entropy, which vanishes linearly
as $T\to 0$, $S_*$ produces a $T$-independent thermal expansion
coefficient $\alpha\propto -\partial S_*/\partial P$ \cite{alpha},
where $P$ is the pressure. In its normal state, CeCoIn$_5$ does in
fact exhibit a greatly enhanced and almost $T$-independent thermal
expansion coefficient \cite{oeschler}, so it is reasonable to
assert that it possesses a flat band.  Analysis of the experimental
data on magnetic oscillations \cite{settai,julian} supports this
assertion. CeCoIn$_5$ is found to have two main Fermi surfaces. The
$\alpha$-sheet is observable at all fields down to $H_{c2}$, while
the magnetic oscillations associated with the $\beta$-sheet become
detectable only at magnetic fields $H\geq 15$ T, behavior in accord
with the posited flat character of this band.

In the theory of fermion condensation, the aforementioned
ground-state degeneracy is lifted at any finite temperature, where
FC acquires a small dispersion proportional to $T$, the spectrum
being given by \cite{noz}
\begin{equation}
\epsilon({\bf p},n_*)=T\ln {1-n_*({\bf p})\over n_*({\bf p})}.
\label{tem}
\end{equation}
However, the lifting of the degeneracy does not change the FC
occupation numbers $n_*({\bf p})$, implying that the entropy excess
$S_*$ would persist down to zero temperature.  To avert a
consequent violation of the Nernst theorem, FC must be completely
eliminated at $T\to 0$. In the most natural scenario, this happens
by means of a SC phase transition, in which FC is destroyed with
the emergence of a pairing gap $\Delta$ in the single-particle
spectrum \cite{ks,physrep,vol2011b,vol2011c,vol2011d}.  We propose
that this scenario is played out in CeCoIn$_5$ at rather weak
magnetic fields $H\ll H_{c2}$, providing for elimination of the
flat portion in the spectrum $\epsilon({\bf p})$ and the removal of
the entropy excess $S_*$.  In stronger external magnetic fields
sufficient to terminate superconductivity in CeCoIn5, this route
becomes ineffective, giving way to an alternative scenario
involving a crossover from FC state to a state having a
multi-connected Fermi surface \cite{shagrep,zb,shagp,prb2008}. In
the phase diagram of CeCoIn$_5$ depicted in Fig.~1, such a
crossover is indicated by the hatched area between the domains of
NFL and LFL behavior and also by the line $T_{\rm cross}(H)$.

We observe that the end point $H_{c0}$ of the curve $T_{\rm
cross}(H)$ nominally separating NFL and LFL phases is a
magnetic-field-induced quantum critical point (QCP) hidden in the
SC state \cite{ronn}.  This the most salient feature of the phase
diagram for the behavior of the resistivity $\rho(T,H)$.  Since the
entropies of the two phases are different, the SC transition must
become first-order \cite{shagrep} near the QCP, in agreement with
the experiment \cite{bianchi}. Moreover, LFL behavior remains in
effect in the domain $T\to 0$, $H>H_{c2}$. It follows that
imposition of fields $H>H_{c2}$ will drive the system from the SC
phase to the LFL phase, where FC, or equivalently the flat portion
of the spectrum $\epsilon({\bf p})$, is destroyed.  Thus,
application of a magnetic field $H>H_{c2}$ to CeCoIn$_5$ is
predicted to cause a step-like drop in its residual resistivity
$\rho_0$, as is in fact seen experimentally \cite{pag1}.
Furthermore, it is to expected that the higher the quality of the
CeCoIn$_5$ single crystal, the greater is the suppression of
$\rho_0$.

\begin{figure}[!ht]
\begin{center}
\includegraphics [width=0.47\textwidth]{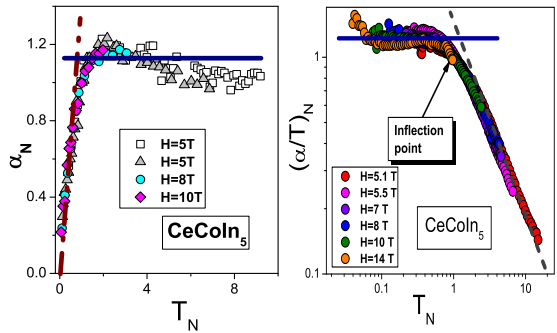}
\end{center}
\caption{(color online). Left panel: Normalized low-temperature
thermal expansion coefficient $\alpha_N$ vs normalized temperature
$T_N$ of the normal state of CeCoIn$_5$ at different magnetic
fields $H$ shown in the legend. All the data represented by the
geometrical symbols are extracted from measurements \cite{steg}.
The dash-dot line indicates the LFL behavior taking place at low
temperatures under the application of magnetic fields. The NFL
behavior at higher temperatures characterized by both $\alpha={\rm
const.}$ and $S_*$ of Eq.~\eqref{S*} is shown as the horizontal
line. Right panel: Normalized low temperature thermal expansion
coefficient $(\alpha/T)_N$ vs $T_N$ at different $H$ shown in the
legend. The data \cite{tompson} and $T$ were normalized by the
values of $\alpha/T$ and by the temperature $T_{inf}\simeq
T_{cross}$, correspondingly, at the inflection point shown by the
arrow. The horizontal solid line depicts the LFL behavior,
$(\alpha/T)={\rm const.}$ The dash line displays the NFL behavior
$(\alpha/T)\propto 1/T$.} \label{fig2}
\end{figure}
As suggested by this analysis of the $H-T$ phase diagram, the
behavior of the dimensionless thermal expansion coefficient,
treated as a function of the dimensionless temperature $T_N$, is
found to be almost universal. Both the left and right panels of
Fig. \ref{fig2} show that all the normalized data extracted from
measurements on CeCoIn$_5$ \cite{steg,tompson} collapse onto a
single scaling curve. As seen from the left panel, the
dimensionless coefficient
$\alpha_N(T,H)=\alpha(T,H)/\alpha(T_N,H)$, treated as a function of
$T_N=T/T_{\rm cross}$, at $T_N<1$ shows a linear dependence, as
depicted by the dash-dot line, implying CeCoIn$_5$ exhibits LFL
behavior in this regime. At $T_N\simeq 1$ the system enters the
narrow crossover region. At $T_N>1$, NFL behavior prevails, and
both $\alpha$ and $S_*$ cease to depend significantly on $T$, with
$\alpha_N$ remaining close to the horizontal line. The observed
limiting behaviors, namely LFL with $\alpha\propto T$ and NFL with
$\alpha={\rm const.}$, are consistent with recent experimental
results \cite{tompson}, as it is seen from the right panel.  From
this evidence we conclude that essential features of the
experimental $T-H$ phase diagram of CeCoIn$_5$ \cite{pag1} are well
represented by Fig. \ref{fig1}.

In calculations of low-temperature transport properties of the normal
state of CeCoIn$_5$, we employ a two-band model, one of which is
supposed to be flat, with the dispersion
given by Eq.~\eqref{tem}, while the second band is assumed to possess
a LFL single-particle spectrum having finite $T$-independent dispersion.
We begin the analysis with the case $H=0$, where the resistivity of
CeCoIn$_5$ is a linear function of $T$.  As will be seen, this behavior
is inherent in electron systems having flat bands.  We first express
the conductivity $\sigma(T)$ in terms of the imaginary part of the
polarization operator $\Pi({\bf j})$ \cite{trio},
\begin{eqnarray} \nonumber
\sigma&=&\lim \omega^{-1}{\rm Im} \Pi({\bf j},\omega\to 0)\propto
{1\over T}\int\int {d\upsilon d\varepsilon \over
\cosh^2(\varepsilon/2T)}\\
&\times&|{\cal T}({\bf j},\omega=0)|^2{\rm Im} G_R({\bf
p},\varepsilon){\rm Im} G_R({\bf p},\varepsilon),\label{con1}
\end{eqnarray}
where $d\upsilon$ is an element of momentum space, ${\cal T}({\bf j},\omega)$
is the vertex part, ${\bf j}$ is the electric current, and
$G_R( {\bf p},\varepsilon)$ is the retarded quasiparticle
Green function, whose imaginary part is given by
\begin{equation}
{\rm Im} G_R({\bf p},\varepsilon)=
-{\gamma\over (\varepsilon-\epsilon({\bf p}))^2+\gamma^2}
\label{gr}
\end{equation}
in terms of the spectrum $\epsilon({\bf p})$ and damping $\gamma$
referring to the band with the finite value $v_F$ of the Fermi
velocity.  Invoking gauge invariance, we have
${\cal T}({\bf j},\omega=0)=e\partial\epsilon({\bf p})/\partial {\bf p}$ .
Upon inserting this equation into Eq.~\eqref{con1} and performing
some algebra we arrive at the standard result
\begin{equation}
\sigma(T)= e^2n{v_F\over \gamma(T)}, \label{con}
\end{equation}
where $n$ is the number density of electrons.

\begin{figure}[!ht]
\begin{center}
\includegraphics [width=0.28\textwidth]{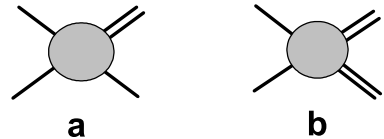}
\end{center}
\caption{Scattering diagrams that contribute to the imaginary part of
the mass operator $\Sigma(\varepsilon)$, referring to the band with
finite value of the Fermi velocity. The single line corresponds to
a quasiparticle of that band, and the double line to a
FC quasiparticle.} \label{fig3}
\end{figure}
In conventional clean metals obeying LFL theory, the damping
$\gamma(T)$ is proportional to $T^2$, leading to Eq.~\eqref{res}
with $n=2$. NFL behavior of $\sigma(T)$ is due to the NFL
temperature dependence of $\gamma(T)$ associated with the presence
of FC \cite{kz,kzc}. In the standard situation where the volume
$\eta$ occupied by FC is rather small, overwhelming contributions
to the transport come from inelastic scattering, represented
diagrammatically in Fig. \ref{fig3}a and Fig. \ref{fig3}b, where FC
quasiparticles (distinguished by the double line) are changed into
normal quasiparticles; or vice versa, normal quasiparticles turn
into FC quasiparticles. Contributions of these processes to the
damping $\gamma$ are estimated on the basis of a simplified formula
\cite{trio}
\begin{equation}
\nonumber
\gamma({\bf p},\varepsilon)\propto\int\int\int\limits_0^\varepsilon
\int\limits_0^\omega |\Gamma({\bf p},{\bf p}_1,{\bf q}|^2{\rm Im}
G_R({\bf p}-{\bf q},\varepsilon-\omega)
\end{equation}
\begin{equation}\times{\rm Im} G_R( -{\bf p}_1,-\epsilon)
{\rm Im} G_R({\bf q}-{\bf p}_1,\omega-\epsilon) d{\bf p}_1 d{\bf q}
d\omega d\epsilon, \label{gamma}
\end{equation}
where now the volume element in momentum space includes
summation over different bands. Calculations whose details can be found in
\cite{kz} yield
\begin{equation}
\gamma(\varepsilon)=\eta(\gamma_0+\gamma_1\varepsilon),\quad {\rm
Re}\Sigma(\varepsilon)=-\eta\gamma_1 \varepsilon \ln {\varepsilon_c\over
|\varepsilon|}, \label{damp}
\end{equation}
with $\eta$ denoting the volume in momentum space occupied by the
flat band, and $\varepsilon_c$ being a characteristic constant,
specifying the logarithmic term in $\Sigma$. Accounting for vertex
corrections \cite{trio} ensures transparent changes in Eq.
\eqref{gamma} and cannot be responsible for the effects discussed
in our article. We note that Eq. \eqref{damp} leads to the lifetime
$\tau_q$ of quasiparticles, $\hbar/\tau_q\simeq a_1+a_2T$, where
$\hbar$ is Planck's constant, $a_1$ and $a_2$ are parameters. This
result is in excellent agreement with experimental observations
\cite{tomph}. Given this result, one finds that
$\rho(T)=\rho_0+AT$, i.e., the resistivity $\rho(T,H=0)$ of systems
hosting FC is indeed a linear function of $T$, in agreement with
experimental data on CeCoIn$_5$. Furthermore, the term $\rho_0$
arises even if the metal has a perfect lattice and no impurities at
all.

The presence of the flat band manifests itself not only in kinetics
but also in the thermodynamics of CeCoIn$_5$, e.g., in the occurrence of
an additional term $\Delta C=C_s-C_n$ in the specific heat $C(T)$,
given by
\begin{equation}
\Delta C=-{1\over 2T}\int \left({d\Delta^2({\bf p})\over
dT}\right)_{T_c}n({\bf p})(1-n({\bf p}))d\upsilon,\label{dC}
\end{equation}
where $\Delta(p)$ is the energy gap and $C_s$ and $C_n$ are the
specific heats of superconducting and normal states, respectively.
It is FC contribution that endows CeCoIn$_5$ with a record value of
the jump $\Delta C/C_n=4.5$\, \cite{petrovic} of the specific heat
at $T_c$ (the LFL value being 1.43). The enhancement factor is
evaluated by setting $T=T_c$ in Eq.~\eqref{dC}. Importantly, in
systems with flat bands, the quantity $q=-(1/2T_c)(d\Delta^2/dT)$
has the same order \cite{noz} as in LFL theory, where $q\simeq 5$.
To illustrate the point, suppose that the momentum distribution
$n({\bf p})$ depends only on the absolute value of ${\bf p}$.  One
then obtains
\begin{equation}
{\Delta C(T_c) \over C_n(T_c)}\sim {v_F\over T_c}\int n( p)
\,\left(1-n( p)\right)dp\,. \label{delc}
\end{equation}
Thus, the ratio $\Delta C(T_c)/C_n(T_c)$, shown to be proportional to the
volume in momentum space occupied by the flat band, behaves as
$1/T_c$, implying that $\Delta C(T_c)/C_n(T_c)$
diverges at $T_c\to 0$, in agreement with data on CeCoIn$_5$, where $T_c$
is only $2.3$ K.

The formula (\ref{dC}) can be recast in the form \cite{yak}
\begin{equation}
{\Delta C(T_c)\over C_n(T_c)}= q{\cal
S}^{-1}(T_c){T_c\chi(T_c)\over C_n(T_c)}, \label{ston}
\end{equation}
suitable for extracting the Stoner factor ${\cal
S}(T_c)=\chi(T_c)/\chi_0(T_c)$. Based on the experimental data
\cite{kim}, one finds ${\cal S}(T_c)\sim 0.3$.

It is straightforward to apply these results to analysis of the
slope of the peak of the specific heat in the superconducting state
of CeCoIn$_5$ as $T\to T_c$, based primarily on Eq.~\eqref{dC}.
More precisely, one has
\begin{equation}  {dC(T\to T_c)/dT\over dC_{BCS}(T\to
T_c)/dT}= {\Delta C \over C_n(T_c)},
\end{equation}
the right side of this relation being evaluated with the aid of Eq.~\eqref{delc}.
This narrowing of the shape of $C(T)$ toward the $\lambda$-point curve,
so familiar in the case of superfluid $^4$He, is in agreement with experimental
data on CeCoIn$_5$.

The component of the damping $\gamma$ linear in energy given by
Eq.~\eqref{damp} is responsible for a logarithmic correction to the
specific heat $C(T)$ of the normal state of CeCoIn$_5$, observed in
\cite{kim}.  With $f(\varepsilon)=[1+e^{\varepsilon/T}]^{-1}$ and
$R(p,\varepsilon) \equiv -i\ln
\left(G_R(p,\varepsilon)/G_A(p,\varepsilon)\right)$, the formula of
\cite{trio} for the entropy $S(T\to 0)$ is recast as
\begin{equation}
S(T)=-T^{-1}\int d\upsilon \int\limits_{-\infty}^{\infty}
\varepsilon  {\partial f(\varepsilon)\over
\partial\varepsilon} R(p,\varepsilon)d\varepsilon.
\end{equation}
where, by virtue of Eq.~(\ref{damp}),
$R(p,\varepsilon) =\tan^{-1}\left(\eta/ \left(1+\eta\ln(\varepsilon_c/
|\varepsilon|) -\epsilon(p)/\varepsilon\right)\right)$.
Changing variables to $w=\epsilon(p)/\varepsilon\propto (p-p_F)/\varepsilon$
and $\varepsilon=zT$ and retaining only leading
terms, $S(T)$ is expressed as the sum $S=S_++S_-$, with
\begin{equation}
S_\pm \propto T\int\limits_0^{\infty} {z^2 e^zdz\over (1+e^z)^2}
\int\limits_{-\infty}^\infty \tan^{-1}
\left[ {\eta\over 1+\eta\ln (\epsilon_c/ T) \mp w}\right]dw.
\end{equation}
These integrals are evaluated analytically to yield
$S(T)-S_{FL}(T)\propto \eta T\ln T$, in agreement with available
experimental data on the specific heat of CeCoIn$_5$ \cite{kim}.

Now we show that the application of field $H>H_{c2}$ on CeCoIn$_5$
generates the step-like drop in the residual resistivity $\rho_0$.
Indeed, as  seen from Fig. \ref{fig1}, at low temperatures
$T<T_{cross}$, the application of fields $H>H_{c2}$ drives the
system from the SC state to the LFL one, where the flat portion of
$\epsilon({\bf p})$ is destroyed. Thus, the term $\eta$ vanishes
strongly reducing $\rho_0$. The suppression of $\rho_0$ in magnetic
fields is associated with another NFL phenomenon recently observed
in CeCoIn$_5$ \cite{julian}. This is the deviation of the
temperature dependence of the amplitude of magnetic oscillations
$A(T)$ from the standard Lifshitz-Kosevich-Dingle form
\begin{equation}
A(T)=A_D \left(X/\sinh X\right).
\label{lk}
\end{equation}
Here $A_D \propto e^{-Y}$ is the Dingle factor, while $Y=2\pi^2k_B
T_D/\omega_c$, $X=2\pi^2k_BT/\omega_c$, $T_D$ is the Dingle
temperature, and $\omega$ is the cyclotron frequency.
Conventionally $T_D$, assumed to be constant, is associated with
impurity scattering.  More generally, $T_D$ is related to the value
of $\rho_0$ in Eq.~(\ref{res}), which, as we have seen, becomes
$T$-{\it dependent} in a system with a flat band.  Why?  Because
the flat band, in the present context, guarantees an overwhelming
contribution to $\rho_0$ that must disappear at $T<T_{\rm
cross}(H)$ to evade violation of the Nernst theorem.  This
inexorable sequence triggers the abrupt downward jump of $\rho_0$
and of $T_D$ correspondingly, leading in turn to an upward jump of
the Dingle factor $A_D$ near $T_{\rm cross}(H)$, which agrees with
observations \cite{julian}. In parallel with this challenging
behavior of the residual resistivity in magnetic fields, it is
apposite to address the behavior of $\rho_0$ versus pressure $P$,
studied in CeCoIn$_5$ experimentally in \cite{sidorov}.  At
$P>P^*=1.6\,{\rm GPa}$, $\rho_0$ drops reversibly by one order of
magnitude to a very small value of about $0.2\,{\rm\mu\Omega cm}$.
We can reasonably infer that the application of pressure eliminates
FC \cite{shag}, triggering a jump in $\rho_0$ to the lower value
measured. It should be emphasized that a nonzero contribution of FC
to $\rho_0$ is associated with the presence of the crystal lattice,
more precisely, with the Umklapp processes, violating momentum
conservation. At the same time, such a restriction is absent in
dealing with the thermal resistivity $w_0$. If, as usual, one
normalizes the thermal resistivity  by $w=\pi^2T/(3e^2\kappa)$
where $\kappa$ is the thermal conductivity, the famous
Wiedemann-Franz relation then reads $\rho_0=w_0$. The distinguished
role of the Umklapp processes in the occurrence of $\rho_0$ in
Fermi systems with FC implies that in the presence of FC, the
Wiedemann-Franz law is violated so that $\rho_0<w_0$. In $\rm
CeCoIn_5$ this violation does take the place \cite{paglione}.

In summary, we have shown that the application of magnetic fields
and pressure on CeCoIn$_5$ leads to strong suppression of the
residual resistivity $\rho_0$. By considering the behavior of the
thermal expansion coefficient, the specific heat, and the amplitude
of magnetic oscillations, we have unveiled the roles played by the
flat band in thermodynamic as well as in transport properties of
CeCoIn$_5$. Our considerations furnish strong evidence for the
presence of flat band in CeCoIn$_5$, which thereby becomes the
member of a long-expected class of Fermi liquids.

We thank A. Alexandrov and Ya. Kopelevich for fruitful discussions.
This work was supported by U.S. DOE, Division of Chemical Sciences,
Office of Basic Energy Sciences, Office of Energy Research, AFOSR
and by the Russian Foundation for Basic Research, as well as the
McDonnell Center for the Space Sciences.

\end{document}